# The Reality of Casimir Friction

**Kimball A. Milton [1],\*, Johan S. Høye [2] and Iver Brevik [3]**

[1] Homer L. Dodge Department of Physics and Astronomy, University of Oklahoma, Norman, OK 73019, USA
[2] Department of Physics, Norwegian University of Science and Technology, Trondheim 7491, Norway; johan.hoye@ntnu.no
[3] Department of Energy and Process Engineering, Norwegian University of Science and Technology, Trondheim 7491, Norway; iver.h.brevik@ntnu.no

\* Correspondence: kmilton@ou.edu; Tel.: +1-405-325-3961 x 36325

**Abstract:** For more than 35 years theorists have studied quantum or Casimir friction, which occurs when two smooth bodies move transversely to each other, experiencing a frictional dissipative force due to quantum electromagnetic fluctuations, which break time-reversal symmetry. These forces are typically very small, unless the bodies are nearly touching, and consequently such effects have never been observed, although lateral Casimir forces have been seen for corrugated surfaces. Partly because of the lack of contact with observations, theoretical predictions for the frictional force between parallel plates, or between a polarizable atom and a metallic plate, have varied widely. Here, we review the history of these calculations, show that theoretical consensus is emerging, and offer some hope that it might be possible to experimentally confirm this phenomenon of dissipative quantum electrodynamics.

## 1. Introduction

The essence of quantum mechanics is fluctuations in sources (currents) and fields. This underlies the great successes of quantum electrodynamics such as explaining the Lamb shift [1] and the anomalous magnetic moment of the electron [2]. Most directly, these fluctuations are manifested through the Casimir effect, the quantum vacuum forces between macroscopic neutral objects. In particular, quantum vacuum fluctuations provide a powerful framework for generalizing van der Waals forces between atoms and molecules. Since its discovery in 1947–1948 [3,4], the Casimir effect has been verified directly in a number of experiments, and has been extended to include repulsive effects [5], induced either by materials or geometry; dynamical effects, the so-called dynamical Casimir effect [6,7], related to Moore-Davies-Fulling-Unruh radiation [8–12]; lateral forces [13], including torques [14,15]; and quantum friction between moving bodies. The latter, which is distinct from either the dynamical Casimir effect or the Unruh effect, has been of intense theoretical interest for almost four decades, but predictions seem to have differed widely. Here we attempt to show that consensus is emerging, and to suggest how observable effects might be seen with an advance of experimental technique. Overviews of the Casimir effect are given, for example, in References [16–19]. Throughout, we use SI (*Système international d'unités*) units.

In this review, we restrict ourselves to friction arising from electromagnetic (photonic) fluctuations, involving motion between polarizable atoms and conducting surfaces. Thus, we do not concern ourselves with other sources of friction, such as that between quantum dots [20] or between graphene sheets [21]. The physics in these cases is rather different, and outside the scope considered. We also do not refer to such effects as Coulomb drag [22,23].

Although the subject of Casimir friction has been a theoretical playground for some decades, because it has not seemed possible to test ideas experimentally, there has not been so much impetus to focus on definitive predictions. The situation bears some resemblance to the early days of quantum electrodynamics; only when it was known from experiment that the Lamb shift and the electron magnetic moment anomaly were small but real, did calculations (which led to the discovery of renormalization) zero in on accurate predictions.

When media are rubbed against each other, frictional forces dissipating energy will be present. This is common daily experience. The evaluation of such forces on a basic level from the fundamental laws of physics is, however, not obvious when energy conservation is present, and the equations of motion on the microscopic level are reversible; *i.e.*, symmetric with respect to the positive and negative time directions. This is the long-standing problem of how the reversible equations of classical mechanics can produce irreversible behavior of thermodynamic processes. A general system of interacting particles left to its own will proceed irreversibly towards thermal equilibrium by which its entropy will always increase towards a maximum at this equilibrium. A central problem is how to show that the reversible equations of mechanics can lead to such an irreversible behavior.

A statistical mechanical way to understand this is that on the microscopic level the number of microstates consistent with a given macrostate has a sharp and very large maximum at equilibrium. The probability of returning to any other macrostate is essentially equal to zero. When media are rubbed against each other, the equilibrium state occurs when their relative velocity has decreased to zero (if allowed to relax). The thermodynamic entropy concept is proportional to the logarithm of the number of microstates, and the steady increase of total entropy shows the irreversible nature of thermodynamics.

Boltzmann studied the problem of irreversible behavior on the microscopic particle level, by considering a gas of interacting particles at low density for which he established his famous Boltzmann equation. This is a combined differential and integral equation taking into account details of two-particle collisions. The gas was described in terms of its particle density as a function of position, velocity, and time. As for the number of binary collisions, statistical averages were used assuming that the colliding particles are uncorrelated. A special feature of this equation, which has classical mechanics as well as Boltzmann's assumption about the rate of collisions (Boltzmann's *Stossanzahlansatz*) as bases, is that it describes an irreversible evolution towards equilibrium. This is the famous *H*-theorem. The quantity *H* is defined similarly to the entropy, but has the opposite sign. It is then straightforward to show that *H* can only decrease with respect to time towards equilibrium. Accordingly, it demonstrates how reversible mechanics gives rise to an irreversible behavior.

The outline of this paper is as follows: In Section 2, we review results known for the frictional force between parallel plates moving transversely to each other. The different methodologies, using quantum statistical mechanics, using quantum field-theoretic methods, and using quantum-mechanical perturbation theory, are described in Section 3. Then the interaction between a single atom and a planar surface is summarized in Section 4. From this the friction between two plates can be inferred as sketched in Section 5. Section 6 briefly describes temperature dependence, which might yield an observable effect. Brief conclusions follow in Section 7.

**2. Friction between Two Plates**

Media rubbed against each other or sliding past each other at short separation form just such a non-trivial problem of irreversible behavior. The standard setup, with which we shall be concerned initially, is shown in Figure 1: there are two parallel dielectric half-spaces (plates of infinite thickness), kept at a fixed small separation *d*. The upper plate is moving with a small (at least nonrelativistic) velocity *v*; the lower plate is at rest. In the simplest versions of the problem, *v* is kept constant. One can allow for the case where the particle density $\varrho_1$ of the lower plate is different from the particle density $\varrho_2$ in the upper plate. This flexibility in formalism makes it possible to deal with the single-particle half-space problem simply by letting $\varrho_2$ approach zero; we will turn to this problem later.

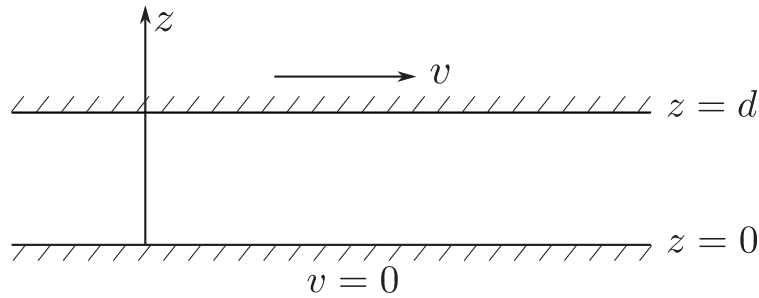

**Figure 1.** Standard configuration: Upper plate, having density ϱ₂, moving with velocity *v*; lower plate, with density ϱ₁, at rest. Gap width is *d*.

It is probably correct to say that the 1978 paper of Teodorovich [24] and the 1989 paper of Levitov [25] started the research on Casimir friction. After that, inspired by the mentioned works two of us contributed also, focusing in the first round on the basic problem of the friction force between a pair of moving harmonic oscillators [26,27]. In subsequent years, there has been a surge of interest from many contributors to the topic.

It is natural to ask what is the reason for the considerably large interest in a phenomenon, which is, after all, rather esoteric. Is the effect of practical importance? The answer so far is definitely no. Let us, as an example, quote from the results of Pendry in his influential papers [28–31]. Assuming constant conductivity σ, and relative permittivity ε = 1 + iσ/(ω ε₀), he derived the following expressions in two limiting cases for the force per unit area *A* [28]:

$$\frac{F}{A} = \frac{5\hbar\varepsilon_0^2 v^3}{2^8 \pi^2 \sigma^2 d^6}, v \ll \frac{d\sigma}{\varepsilon_0} \quad (1)$$

$$\frac{F}{A} = \frac{\hbar\sigma^2}{32 d^2 \pi^2 v \varepsilon_0^2} \ln\frac{v\varepsilon_0}{2d\sigma}, v \gg \frac{d\sigma}{\varepsilon_0} \quad (2)$$

(Here, $\varepsilon_0 = 8.854 \times 10^{-12}$ F/m is the permittivity of free space.) Now assuming the effective gap *d* to be 1 nm, and estimating σ for a semiconductor to be about 0.1 S/m, we obtain with a moderate velocity of $v = 1.0$ m/s that $F/A \cong 1.6$ mPa. Even this small number corresponds to an unrealistically small separation; and for metal plates the frictional force becomes really tiny, $F/A = 8\times 10^{-21}$ Pa for $d = 1$ nm and $\sigma = 4.5 \times 10^7$ S/m as appropriate for gold. This is particularly insignificant as compared to the conventional Casimir force between ideal metal plates at rest at 1 nm separation, $F_C = 1.3 \times 10^9$ Pa. (At such distances, a more apt comparison is to consider the non-retarded van der Waals force, which becomes roughly smaller by a factor of $d/\lambda$, λ being a wavelength characteristic of the atomic polarizability. In fact, we estimate the van der Waals force to be [32,33] $F_{vdW}/A \leq 1 \times 10^7$ Pa). One would think that increasing the velocity could enhance the effect. For the case of gold, even for $v = c$ (where the inequality in Equation (1) above is still satisfied), with the same parameters as above, we get $F/A = 2 \times 10^5$ Pa, which is still small, even for this very unrealistic extreme case, compared to the conventional Casimir or van der Waals pressure. (Of course, the formulas are not valid for relativistic velocities.) For the case of dielectric half spaces with $\sigma = 0.1$ S/m, $d = 1$ nm, the two regimes in Equations (1) and (2) cross at $v = 10$ m/s, and there the approximate maximum frictional force per unit area only reaches 2 Pa. For larger velocities, the force gets weaker according to Pendry's second Equation (2) above.

Thus, the significant interest in these phenomena stems not from their practical usefulness but rather from the fact that this is a relatively simple example of the interplay between time-reversible mechanics/electrodynamics and thermodynamics.

From an optical point of view it is of interest to trace out how the Casimir friction compares with the Doppler effect for photons. Assume that there are two half-planes moving longitudinally with respect to each other, and let an emitter at rest in the right-moving upper plane send out a photon that is received by an absorber at rest in the lower plane. When the photon is moving to the right it will be Doppler shifted to a higher frequency and thus a higher momentum is transferred to the lower plate at rest when the photon is absorbed by it. Likewise when the photon is moving to the left it will be Doppler shifted to a lower frequency, and a corresponding lower momentum in the opposite direction is transferred by absorption. This imbalance in momentum transfer gives the Casimir friction. This is, for instance, illustrated in Figures 2 and 4 in References [28] and [34], respectively.

The frictional force, in the ordinary sense considered here, is due to the dissipation of energy. There a key element is absorption, represented by the imaginary part $\varepsilon''$ of the permittivity (assuming a nonmagnetic medium). We may recall that for a monochromatic wave in a dispersive medium the heat developed per unit time and volume is equal to $\omega \varepsilon''(\omega)$ times the spectral electromagnetic energy density; *cf.*, for instance, Reference [35]. (This leaves out special cases such as Cherenkov radiation, where the superluminal condition $v > c/n$, $n$ being the index of refraction, permits elementary waves emitted from a uniformly moving particle to accumulate on a cone, called the Cherenkov cone. The Cherenkov radiation is obviously also associated with a braking force.)

As a side remark, we observe that the same requirement of dissipation is encountered also in the standard case of a normal Casimir force between the two planes when evaluated via the common method of fluctuating electric fields [36,37]. The central quantity in this context is the two-point function—the expectation value of the product of electric fields at two spacetime points. For a given temperature this quantity turns out to be proportional to the imaginary value of the retarded Green function in Fourier (frequency-wavevector) space; this being the fluctuation-dissipation theorem. Unless the medium has some absorption, the imaginary part of this function is simply zero. This point is usually hidden somewhat by the formalism, since the frequency integration is commonly transformed into an integration over imaginary frequencies, the so-called Wick, or more properly, Euclidean, rotation.

Another aspect of Casimir friction of interest in optics is the appearance of the anomalous Doppler effect, where the shifted frequency is negative. This is usually associated with the superluminal condition $v > c/n$ mentioned above. A detailed account of this effect can be found in the volume by Ginzburg [38], and also in Reference [39]. The effect has remarkably enough an analogy also in the present case: Assume that a two-level atom moves parallel to a dielectric surface with low velocity $v$. Assume that the atom has a Bohr transition frequency $\Omega$ between the two states. The resonance condition for radiation is that $\Omega + \omega' = 0$, where $\omega'$ is the photon frequency in the atom's comoving frame. Now, because of the Doppler effect, $\omega' = \omega - k \cdot v$, where $k$ is the wave vector of the photon ($\omega$ is the emitted frequency in the laboratory frame). A spontaneous excitation of the atom's ground state becomes possible. This is the point being analogous to the anomalous Doppler effect in the superluminal case. This brief argument is explored in more detail in References [40,41]. We mention also that an analysis of a quantum point detector uniformly accelerated through a uniform dielectric medium can be found in Reference [42].

We note in passing that the conventional superluminal anomalous Doppler effect gives a striking example of the usefulness of the Minkowski energy-momentum tensor. Since this tensor is divergence-free, it causes the total energy and the total momentum of a radiation field in a medium to form a four-vector. The most characteristic feature of this four-vector is that it is space-like, thus permitting a negative field energy in certain inertial frames. Experimentally, this peculiar property is manifested clearly when considering the Cherenkov effect in the rest system of the emitter: the recoil of the particle gives a positive energy, thus the energy of the photon has to be negative. Some more discussion on this point can be found in Reference [43]. See also References [44,45].

### 3. Methodologies

When categorizing the various approaches to the Casimir friction problem, one may roughly distinguish between the following cases:

One way is to apply methods from *quantum statistical mechanics* to the system of relatively moving harmonic oscillators, in general at a finite temperature $T$. This is the method that two of us have used repeatedly in previous recent investigations [46–53]; *cf.* also the earlier papers [26,27] where the basis of the formalism was laid. The method is quite compact and effective. The essence of it is to generalize the statistical mechanical Kubo formalism [54], which can describe non-equilibrium phenomena, to time-dependent cases. The polarizable particles are considered as harmonic oscillators where the dipole moments fluctuate (oscillate), and pairs of oscillators interact via the electromagnetic dipole interaction. Methods of classical statistical mechanics of fluids had earlier been extended to polarizable particles where the harmonic oscillations were quantized [55,56]. (This, of course, is a fundamental approach.) This extension was possible via the Feynman path integral representation where quantized particles can be regarded as "classical" polymers (or random walks) that extend in a fourth dimension, imaginary time, of length $\hbar c \beta = \hbar c/(k_B T)$, where $k_B$ is Boltzmann's constant, and $T$ is temperature. In this picture, Casimir forces can be interpreted as induced interactions due to quantized fluctuating dipole moments that interact via the radiating (time-dependent) dipole-dipole interaction. Within the "polymer" picture, arguments and methods of classical statistical mechanics can be applied to obtain Fourier transforms of time-dependent response functions given by quantum mechanical commutators. They follow from the corresponding correlation functions of the classical problem in imaginary time. It turns out that the viewpoint with fluctuating dipole moments is equivalent to the more traditional one where medium-induced changes in the ground state energies of the quantized electromagnetic field induce the Casimir forces.

A second method is more conventional, namely to apply *quantum field theoretical methods*. Already in the mentioned Levitov paper [25] the van der Waals (Casimir) friction between dielectric slabs moving past each other at close separation was considered. Polarization currents in each of the two bodies interact via the electromagnetic field. Photons transferred between the slabs suffer Doppler shifts due to the relative motion. The recent series of articles by Volokitin and Persson are following the same kind of approach [57–61] as do Dedkov and Kyasov [62–66]. The work of Polevoi [67] and Mkrtchian [68] should be mentioned. The fluctuation-dissipation theorem, which is equivalent to the Kubo formalism, here plays an important role. The authors derive a field theoretical formalism leading to the power spectral density of the fluctuating electromagnetic field, and apply it to the radiative heat transfer and the Casimir friction both using the semiclassical theory of fluctuating fields [69,36] and the full quantum theory. Especially Reference [34] (a review) contains a great deal of information.

Third, a special class of approaches is to make use of *quantum mechanical perturbation theory*. The investigations of Barton [70–74] may be said to belong to this class. Both microscopic harmonic oscillators, and half-spaces, were studied. A recent discussion, which modifies some of Barton's results, is Reference [40]. We could also regard the classic result of Pendry [28] to fall in this category.

For the methods mentioned above, widely different physical mechanisms are considered as basis for the friction force. With the quantum statistical method a system at thermal equilibrium is perturbed by a time-dependent interaction due to the relative speed, and the Kubo formalism is utilized. With this method one evaluates a response function of the system. This function gives the influence on a property of interest of the system when acted upon by a perturbing interaction. With this the friction force is obtained either directly or more reliably via the energy dissipated. By the second method, applicable to electromagnetic interactions, the well-known Doppler effect for photons emitted and scattered from moving media is used as basis. Finally the third method uses time-dependent perturbation theory of quantized systems as the physical basis. Then the energy dissipated (absorbed) due to excitations and deexcitations of the system is evaluated. With such widely different viewpoints there can be good reason to ask whether they all lead to consistent results in the end. An explicit demonstration of the equivalence between the Høye/Brevik results and those obtained by Barton was given in Reference [48].

There are a large number of other papers on Casimir friction, for example, valuable work given in References [75–81], looking at the problem from different angles. (Philbin and Leonhardt [75] obtain no friction, but we believe their analysis is flawed.) For example, Nesterenko and Nesterenko [80] obtain the same velocity dependence given below, proportional to the cube of the velocity, but are unable to determine the dependence on separation. The list of references is not exhaustive, although to our knowledge it covers the majority of the recent contributions. As the Casimir friction concept is a delicate one where theoretical predictions obtained in the literature are sometimes wildly diverging, it is of interest to trace out how far the development has come in getting results in agreement with each other. In the following we sketch some of the results found and how they compare.

## 4. Interaction between an Atom and a Plate

The basic problem of the friction force on an atom moving through the thermal radiation field in vacuum goes back to Einstein and Hopf [82,83]; see also Reference [84]. Beyond that it is perhaps easiest to start by considering a single polarizable atom that also moves through the thermal radiation field in vacuum, but in addition it interacts with a planar substrate, say an imperfect metal, as shown in Figure 2. In order to have friction, both bodies must possess dissipation. The dissipation of the atom may be due either to the intrinsic radiation reaction of the electromagnetic field, or due to the dissipation in the metal surface in which the atom is reflected. Radiation reaction is an inescapable effect, but usually very small compared to dissipation within metals, as follows from the results given below.

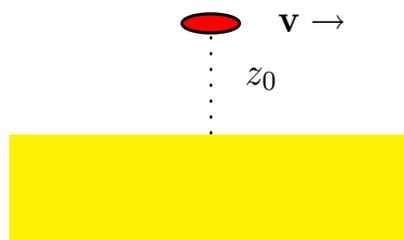

**Figure 2.** Atom moving above a dielectric or metallic surface. Although the medium and the atom may be taken to be isotropic, the electromagnetic interaction induces anisotropy. See, for example Reference [85].

If the surface is metallic, the dissipation in the bulk is provided by the resistivity of the metal, or its conductivity $\sigma$. The dominant loss mechanism of the atom is due to the damping also provided by the surface; this can be thought of as the dissipation of the image of the atom moving through the bulk. (The interactions between the bodies can be described in terms of multiple reflections, which captures the dissipative effect of the bulk. See also Reference [86].) Then using either the Kubo formalism or the fluctuation-dissipation theorem we find for low velocities that the frictional force is [85]

$$F = \frac{135\hbar\alpha^2 v^3}{4\pi^3\sigma^2(2z_0)^{10}} \qquad (3)$$

where $\alpha$ is the static polarizability of the atom (assumed isotropic), $v$ is its velocity parallel to the surface, and $z_0$ is the distance between the atom and the surface. This result agrees within a factor of 3/8 with that found in Reference [81]. The salient dependence is upon the cube of the velocity and the inverse tenth power of the distance. This friction can become appreciable only if the atom is extremely close to the surface. It is noteworthy that the same result (within a further factor of 5) was obtained by a perturbative method, related to that of Barton [71] in a recent paper [40], which argues that the linear velocity dependence found by Barton is an artifact of the particular velocity profile assumed by him. Scheel and Buhmann [87] like Barton had also obtained such a linear dependence. See also Reference [88]. (Some of the linear effects found reflect interactions with excited states, rather than with ground-state atoms, which is our focus here.) Another example of congruence of results appears in a recent paper [89] where a formula (Formula (41)) is given for the frictional force experienced by a dielectric particle moving in a thermal field, the problem considered by Einstein and Hopf [82,83]. This formula is exactly the same (apart from notation) as that derived in Equations (A4) and (A15) in Reference [27] and is given by:

$$F = \frac{\beta\hbar^2\ \alpha\omega^6 v}{6c^5\ \sinh^2(\beta\hbar\omega/2)} \quad (4)$$

where ω is the resonance frequency, $\omega^2 = e^2/(m\ \alpha)$, for an oscillating charge $e$ with mass $m$, and $\beta = 1/k_B T$.

Numerically, as an example using the numbers appropriate to a rubidium atom moving past a silicon surface at the speed of sound, perhaps relevant to the atomic source, $v = 340$ m/s, at a distance of $z_0 = 10$ nm, where the parameters characterizing the atom and the surface are $\alpha = 4\pi\varepsilon_0 \times 4.7 \times 10^{-29}$ m$^3$, $\sigma = 0.0015$ S/m, the frictional force (3) on the atom is $F = 5 \times 10^{-21}$ N. In spite of the optimism expressed in Reference [81], this is several orders of magnitude smaller than the conventional van der Waals or Casimir-Polder force between the atom and the substrate, the former being given by [17]:

$$F_{vdW} = \frac{3\hbar c\alpha}{16\pi\varepsilon_0 \lambda_0 z_0^4} \sim 10^{-14}\ \text{N} \quad (5)$$

where $\lambda_0$ is the characteristic absorption wavelength.

The above dependence of the frictional force critically depends on the dissipative mechanism assumed. In the vacuum radiation field of empty space (without the presence of the plate), the atom would only suffer dissipation due to radiation reaction. The latter is due to emission and absorption of dipole radiation to obtain equilibrium with its surroundings. Due to this the oscillations of the atomic polarization are damped which, together with the conductivity σ of the metal plate, again leads to dissipation. If this were the dominant mechanism, the same calculation referred to above with the plate at distance $z_0$ would result in a quite different frictional force dependence [85]:

$$F = \frac{105\hbar\alpha^2 v^5}{32\pi c^3 \varepsilon_0 \sigma z_0^9} \quad (6)$$

slightly weaker distance dependence, and now going as the 5th power of the velocity.

This frictional effect is always subdominant to the induced effect described above, being 20 orders of magnitude smaller for the parameters given above. The change in power in velocity is due to different powers of the frequency ω for the imaginary part of the resulting effective polarizability. Only small ω → 0 are relevant for small velocities. In the two cases the imaginary parts are proportional to ω and $\omega^3$ respectively. The former is influenced by the resistivity of the metal plane. Thus the corresponding friction force (3) contains two factors of σ.

It therefore appears that (perhaps up to a polarization factor) congruence between different calculations for the force between an atom and a surface has been achieved, but the conclusion is inescapable: the force seems far beyond experimental reach.

## 5. Friction between Parallel Plates

After this discussion of atom-surface friction, we return to the force between two relatively moving surfaces. We can obtain this, in fact, by considering the force between an atom and a half space, and then considering a dilute gas of such atoms. Of course, Casimir forces are not additive, so it is nontrivial to generalize to the interaction between two dense bodies. One can replace the polarization by the permittivity by using the substitution derived in Reference [51] (justified because the low frequency regime dominates—see also, for example, Reference [17])

$$\frac{\rho\alpha}{2\varepsilon_0} \to \frac{\varepsilon - \varepsilon_0}{\varepsilon + \varepsilon_0} \quad (7)$$

where the right-hand side is the transverse magnetic reflection coefficient at zero frequency. Using the Kubo formula, two of us have obtained [52] at zero temperature the result of Pendry given above multiplied by a factor of 12 [52]

$$\frac{F}{A} = \frac{15\hbar\varepsilon_0^2 v^3}{64\pi^2 \sigma^2 d^6} \tag{8}$$

This result agrees with that of Barton [73] (apart from a factor of ζ(5) = 1.037, presumably due to the neglect of multiple scattering), and differs by a factor of 2 from that of Volokitin and Persson [34]:

$$F_{HB} = F_B = 2F_{VP} = 12F_P \tag{9}$$

The factor 2 is presumably due to integral (87) in Reference [34] where the integration over the wave vector $q_x$ is restricted to $q_x > 0$ while integral (48) in Reference [52] does not have this restriction. (The factor of 6 between Pendry's [28] and Volokitin and Persson's result seems to be simply a miscalculation by Pendry.) We believe that essential convergence of results has thus been achieved.

An obvious extension is to consider the friction of a rotating body near a stationary plate. A recent calculation of a rotating particle above a heated surface find s a very small effect [90]. There are a number of other recent works on the subject of rotational friction, for example Reference [91].

## 6. Temperature Dependence

Nearly all of the above refers to zero temperature. Calculations have also been carried out for finite temperature. For example, Høye and Brevik [51,52] find at "high" temperature ($k_B T \gg \frac{\hbar v}{d}$, *i.e.*, unless $T$ is very small).

$$\frac{F_T}{F_0} = \frac{16\pi^2}{15}\left(\frac{dk_B T}{\hbar v}\right)^2 \tag{10}$$

$F_0 = F$ being the zero-temperature force given in Equation (8). This formula (apart from a factor 1.2 probably due to neglect of multiple scattering) is in agreement with that of Volokitin and Persson [34]. (The reason that this high-temperature limit is nonclassical is that, like the total energy contained in the Planck black-body spectrum, it involves an integral over all frequencies, cutoff by the thermal frequency.) For room temperature, and $d$ = 1nm and a relative velocity $v$ = 1 m/s Equation (10) becomes a very significant enhancement factor, $F_T/F_0 = 1.5 \times 10^{10}$, so this suggests the thermal Casimir frictional force might well be observable. Of course, this is an unrealistically small separation, so without clever techniques to measure the frictional force, perhaps optical, observing Casimir friction can, so far, only remain a dream for the future.

## 7. Conclusions

Although the subject of Casimir friction has been discussed for nearly four decades, it has remained a theoretical playground, and perhaps as a consequence, only now are convergent predictions emerging. Physics is an experimental science, so it is imperative that observations be brought to bear on this phenomenon, which lies at the intersection of quantum mechanics and nonequilibrium statistical mechanics. In this brief review, we acknowledge that such experiments are highly challenging, since the atoms and plates must be brought very close together to get significant effects, which tend to be dominated by other forces, such as the ordinary static (nonretarded) Casimir force. Nevertheless, it seems that the thermal frictional effect might be accessible to experimental study.

**Acknowledgments:** Kimball A. Milton thanks the Julian Schwinger Foundation for partial support of this research. He thanks Diego Dalvit, Philip Hwang, Taylor Murphy, and Prachi Parashar for helpful comments.

**Author Contributions:** All three authors contributed equally to the writing of this review.

**Conflicts of Interest:** The authors declare no conflict of interest.

## Abbreviations

The following abbreviations are used in this manuscript:

| | |
|---|---|
| SI | Système international d'unités |
| TM | Transverse magnetic |
| C | Casimir |
| CP | Casimir-Polder |
| vdW | van der Waals |

**References**

1. Mohr, P.J.; Taylor, B.N.; Newell, D.B. CODATA Recommended Values of the Fundamental Physical Constants: 2006. *Rev. Mod. Phys.* **2008**, *80*, 633–730.
2. Hanneke, D.; Fogwell, S.; Gabrielse, G. New Measurement of the Electron Magnetic Moment and the Fine Structure Constant. *Phys. Rev. Lett.* **2008**, *100*, 120801, doi:10.1103/PhysRevLett.100.120801.
3. Casimir, H.B.G.; Polder, D. The Influence of retardation on the London-van der Waals forces. *Phys. Rev.* **1948**, *73*, 360, doi:10.1103/PhysRev.73.360.
4. Casimir, H.B.G. On the Attraction Between Two Perfectly Conducting Plates. *Kon. Ned. Akad. Wetensch. Proc.* **1948**, *51*, 793–795.
5. Milton, K.A.; Abalo, E.K.; Parashar, P.; Pourtolami, N.; Brevik, I.; Ellingsen, S.Å. Repulsive Casimir and Casimir-Polder Forces. *J. Phys. A* **2012**, *45*, 374006, doi:10.1088/1751-8113/45/37/374006.
6. Wilson, C.M.; Johansson, G.; Pourkabirian, A.; Simoen, M.; Johansson, J.R.; Duty, T.; Nori, F.; Delsing, P. Observation of the dynamical Casimir effect in a superconducting circuit. *Nature* **2011**, *479*, 376–379.
7. Lähteenmäkia, P.; Paraoanua, G.S.; Hasselb, J.M; Hakonena, P.J. Dynamical Casimir effect in a Josephson metamaterial. *PNAS* **2013**, *110*, 4234–4238.
8. Moore, G. T. Quantum theory of the electromagnetic field in a variable-length one-dimensional cavity. *J. Math. Phys.* **1970**, *11*, 2679.
9. Fullling, S.A. Nonuniqueness of Canonical Field Quantization in Riemannian Space-Time. *Phys. Rev. D* **1973**, *7*, 2850, doi:10.1103/PhysRevD.7.2850.
10. Davies, P.C.W. Scalar production in Schwarzschild and Rindler metrics. *J. Phys. A* **1975**, *8*, 609, doi:10.1088/0305-4470/8/4/022.
11. Unruh, W.G. Notes on black-hole evaporation. *Phys. Rev. D* **1976**, *14*, 870, doi:10.1103/PhysRevD.14.870.
12. Fulling, S.A.; Matsas, G.E.A. Unruh Effect. *Scholarpedia* **2014**, *9*, 31789, doi:10.4249/scholarpedia.31789.
13. Chen, F.; Mohideen, U.; Klimchitskaya, G.L.; Mostepanenko, V.M. Experimental and theoretical investigation of the lateral Casimir force between corrugated surfaces. *Phys. Rev. A*, **2002**, *66*, 032113, doi:10.1103/PhysRevA.66.032113.
14. Munday, J.M.; Iannuzzi, D.; Barash, Y.; Capasso, F. Torque on birefringent plates induced by quantum fluctuations. *Phys. Rev. A* **2005**, *71*, 042102, doi:10.1103/PhysRevA.71.
15. Guérout, R.; Genet, C.; Lambrecht, A.; Reynaud, S. Casimir Torque between Nanostructured Plate. *EPL* **2015**, *111*, 44001, doi:10.1209/0295-5075/111/44001.
16. Milton, K.A. *The Casimir Effect: Physical Manifestations of Zero-Point Energy*; World Scientific: Singapore, 2001.
17. Bordag, M.; Klimchitskaya, G.I.; Mohideen, U.; Mostepanenko, V.M. *Advances in the Casimir Effect*; Oxford University Press: Oxford, UK, 2009.
18. Dalvit, D.A.R.; Milonni, P.; Roberts, D.; da Rosa, F. (Eds.) *Casimir Physics*; Springer: Berlin, Germany, 2011.
19. Simpson, W.M.R.; Leonhardt, U. *Force of the Quantum Vacuum: An Introduction to Casimir Physics*; World Scientific: Singapore, 2015.
20. Mate, C.M.; McClelland, G.M.; Erlandsson, R.; Chiang, S. Atomic-Scale Friction of a Tungston Tip on a Graphite Surface. *Phys. Rev. Lett.* **1987**, *59*, 1942, doi:10.1103/PhysRevLett.59.1942.
21. Berman, D.; Erdemir, A.; Zinovev, A.V.; Sumant, A.V. Nanoscale friction properties of graphene and graphene oxide. *Diam. Rel. Mater.* **2015**, *54*, 91–96.
22. Levchenko, A.; Kamenev, A. Coulomb Drag at Zero Temperature. *Phys. Rev. Lett.* **2008**, *100*, 026805, doi:10.1103/PhysRevLett.100.026805.
23. Persson, B.N.J.; Zhang, Z. Theory of friction: Coulomb drag between two closely spaced solids. *Phys. Rev. B* **1998**, *57*, 7327, doi:10.1103/PhysRevB.57.7327.

24. Teodorovich, E.V. Contribution of macroscopic van der Waals interactions to frictional force. *Proc. R. Soc. Lond. A* **1978**, *362*, 71–77.
25. Levitov, L.S. Van der Waals friction. *Europhys. Lett.* **1989**, *8*, 499–504.
26. Høye, J.S.; Brevik, I. Friction force between moving harmonic oscillators. *Physica A* **1992**, *181*, 413–426.
27. Høye, J.S.; Brevik, I. Friction force with non-instantaneous interaction between moving harmonic oscillators. *Physica A* **1993**, *196*, 241–254.
28. Pendry, J.B. Shearing the vacuum—Quantum friction. *J. Phys. Condens. Matter* **1997**, *9*, 10301–10320.
29. Pendry, J.B. Can sheared surfaces emit light? *J. Mod. Opt.* **1998**, *45*, 2389–2408.
30. Pendry, J.B. Quantum friction—Fact or fiction? *New J. Phys.* **2010**, *12*, 033028, doi:10.1088/1367-2630/12/3/033028.
31. Pendry, J.B. Reply to comment on "Quantum friction—Fact or friction?" *New J. Phys.* **2010**, *12*, 068002, doi:10.1088/1367-2630/12/6/068001.
32. Landau, L.D.; Lifshitz, E.M. *Statistical Physics, Part 2*; Pergamon Press: Oxford, UK, 1980.
33. Lambrecht, A.; Reynaud, S. Casimir force between metallic mirrors. *Eur. Phys. J. D* **2000**, *8*, 309–318.
34. Volokitin, A.I.; Persson, B.N.J. Near-field radiative heat transfer and noncontact friction. *Rev. Mod. Phys.* **2007**, *79*, 1291–1329.
35. Landau, L.D.; Lifshitz, E.M. *Electrodynamics of Continuous Media*; Pergamon Press: Oxford, UK, 1984.
36. Rytov, S.M. *Theory of Electrical Fluctuations and Thermal Radiation*; Academy of Science USSR Publishing House: Moscow, Russia, 1953.
37. Milonni, P.W. *The Quantum Vacuum: An Introduction to Quantum Electrodynamics*; Academic Press: Cambridge, MA, USA, 1994.
38. Ginzburg, V.L. *Applications of Electrodynamics in Theoretical Physics and Astrophysics*; Gordon and Breach Science Publisher: Philadelphia, PA, USA, 1989.
39. Ginzburg, V.L. Radiation by uniformly moving sources (Vavilov-Cherenkov effect, transition radiation, and other phenomena). *Phys. Uspekhi* **1996**, *39*, 973–982.
40. Intravaia, F.; Mkrtchian, V.E.; Buhmann, S.; Schell, S.; Dalvit, D.A.R.; Henkel, C. Friction forces on atoms after acceleration. *J. Phys. Condens. Matter* **2015**, *27*, 214020, doi:10.1088/0953-8984/27/21/214020.
41. Pieplow, G.; Henkel, C. Cherenkov friction on a neutral particle moving parallel to a dielectric. *J. Phys. Condens. Matter* **2015**, *27*, 214001, doi:10.1088/0953-8984/27/21/214001.
42. Brevik, I.; Kolbenstvedt, H. Quantum point detector moving through a dielectric medium. II. Constant acceleration. *Il Nuovo Cimento* B **1989**, *103*, 45–62.
43. Brevik, I.; Lautrup, B. Quantum Electrodynamics in Material Media; Munksgaard: Copenhagen, Denmark, 1970; Volume 38, pp. 1–37.
44. Barnett, S.M. Resolution of the Abraham-Minkowski Dilemma. *Phys. Rev. Lett.* **2010**, *104*, 070401, doi:10.1103/PhysRevLett.104.070401.
45. Wang, S.; Ng, J.; Xiao, M.; Chan, C.T. Electromagnetic stress at the boundary: Photon pressure or tension? *Sci. Adv.* **2016**, *2*, e1501485.
46. Høye, J.S.; Brevik, I. Casimir friction force and energy dissipation for moving harmonic oscillators. *EPL* **2010**, *91*, 60003, doi:10.1209/0295-5075/91/60003.
47. Høye, J.S.; Brevik, I. Casimir friction force and energy dissipation for moving harmonic oscillators. II. *Eur. Phys. J. D* **2011**, *61*, 335–339.
48. Høye, J.S.; Brevik, I. Casimir friction in terms of moving harmonic oscillators: Equivalence between two different formulations. *Eur. Phys. J. D* **2011**, *64*, 1–3.
49. Høye, J.S.; Brevik, I. Casimir friction force between polarizable media. *Eur. Phys. J. D* **2012**, *66*, 149.
50. Høye, J.S.; Brevik, I. Casimir friction force for moving harmonic oscillators. *Int. J. Mod. Phys. A* **2012**, *27*, 1260011, doi:10.1142/S0217751X12600111.
51. Høye, J.S.; Brevik, I. Casimir friction between dense polarizable media. *Entropy* **2013**, *15*, 3045–3064.
52. Høye, J.S.; Brevik, I. Casimir friction at zero and finite temperatures. *Eur. Phys. J. D* **2014**, *68*, 61.
53. Høye, J.S.; Brevik, I. Casimir friction: Relative motion more generally. *J. Phys. Condens. Matter* **2015**, *27*, 214008, doi:10.1088/0953-8984/27/21/214008.
54. Kubo, R. In *Lectures in Theoretical Physics*; Brittin, W.E., Dunham, L.G., Eds.; Interscience: New York, NY, USA, 1959; Volume 1.
55. Høye, J.S.; Stell, G. Quantum statistical mechanical model for polarizable fluids. *J. Chem. Phys.* **1981**, *75*, 5133.

56. Thompson, M.J.; Schweizer, K.S.; Chandler, D. Quantum theory of polarization in liquids: Exact solution of the mean spherical and related approximations. *J. Chem. Phys.* **1982**, *76*, 1128–1135.
57. Volokitin, A.I.; Persson, B.N.J. Theory of friction: The contribution from a fluctuating electromagnetic field. *J. Phys. Condens. Matter* **1999**, *11*, 345–359.
58. Volokitin, A.I.; Persson, B.N.J. Noncontact friction between nanostructures. *Phys. Rev. B* **2003**, *68*, 155420, doi:10.1103/PhysRevB.68.155420.
59. Volokitin, A.I.; Persson, B.N.J. Theory of the interaction forces and the radiative heat transfer between moving bodies. *Phys. Rev. B* **2008**, *78*, 155437, doi:10.1103/PhysRevB.78.155437.
60. Volokitin, A.I.; Persson, B.N.J. Quantum friction. *Phys. Rev. Lett.* **2011**, *106*, 094502, doi:10.1103/PhysRevLett.106.094502.
61. Volokitin, A.I.; Persson, B.N.J. Comment on "Fully covariant radiation force on a polarizable particle". *New J. Phys.* **2014**, *16*, 118001, doi:10.1088/1367-2630/16/11/118001.
62. Dedkov, G.V.; Kyasov, A.A. Vacuum attraction, friction and heating of nanoparticles moving nearby a heated surface. *J. Phys. Condens. Matter* **2008**, *20*, 354006, doi:10.1088/0953-8984/20/35/354006.
63. Dedkov, G.V.; Kyasov, A.A. Conservative-dissipative forces and heating mediated by fluctuating electromagnetic field: Two plates in relative nonrelativistic motion. *Surf. Sci.* **2010**, *604*, 562–567.
64. Dedkov, G.V.; Kyasov, A.A. Dynamical van der Waals atom-surface interaction. *Surf. Sci.* **2011**, *605*, 1077–1081.
65. Dedkov, G.V.; Kyasov, A.A. Dynamical Casimir-Polder atom-surface interaction. *Surf. Sci.* **2012**, *606*, 46–52.
66. Dedkov, G.V.; Kyasov, A.A. A uniformly moving and rotating polarizable particle in thermal radiation field: Frictional force and torque, radiation and heating. 2015, arXiv:1504.01588.
67. Polevoi, V.G. Tangential molecular forces between moving bodies by a fluctuating electromagnetic field. *Sov. Phys. JETP* **1990**, *71*, 1119.
68. Mkrtchian, V.E. Interaction between moving macroscopic bodies: Viscosity of the electromagnetic vacuum. *Phys. Lett. A* **1995**, *207*, 299–302.
69. Rytov, S.M.; Kravtsov, Y.A. Elements of Random Fields. In *Principles of Statistical Radiophysics*; Springer: Berlin, Germany, 1989; Volume 3.
70. Barton, G. On van der Waals friction: I. Between two atoms. *New J. Phys.* **2010**, *12*, 113044, doi:10.1088/1367-2630/12/11/113044.
71. Barton, G. On van der Waals friction. II: Between atom and half-space. *New J. Phys.* **2010**, *12*, 113045, doi:10.1088/1367-2630/12/11/113045.
72. Barton, G. On van der Waals friction between two atoms at nonzero temperature. *New J. Phys.* **2011**, *13*, 043023, doi:10.1088/1367-2630/13/4/043023.
73. Barton, G. On van der Waals friction between half-spaces at low temperature. *J. Phys. Condens. Matter* **2011**, *23*, 335004, doi:10.1088/0953-8984/23/35/355004.
74. Barton, G. Van der Waals friction: A Hamiltonian test bed. *Int. J. Mod. Phys. A* **2012**, *27*, 1260002, doi:10.1142/S0217751X12600020.
75. Philbin, T.G.; Leonhardt, U. No quantum friction between uniformly moving plates. *New J. Phys.* **2009**, *11*, 033035, doi:10.1088/1367-2630/11/3/033035.
76. Pieplow, G.; Henkel, C. Fully covariant radiation force on a polarizable particle. *New J. Phys.* **2013**, *15*, 023027, doi:10.1088/1367-2630/15/2/023027.
77. Maghrebi, M.F.; Golestanian, R.; Kardar, M. Quantum Cherenkov radiation and noncontact friction. *Phys. Rev. A* **2013**, *88*, 042509, doi:10.1103/PhysRevA.88.042509.
78. Silveirinha, M.G. Theory of quantum friction. *New J. Phys.* **2014**, *16*, 063011, doi:10.1088/1367-2630/16/6/063011.
79. Fröhlich, J.; Gang, Z. Emission of Cherenkov Radiation as a Mechanism for Hamiltonian Friction. *Adv. Math.* **2014**, *264*, 183–235.
80. Nesterenko, V.V.; Nesterenko, A. V. Macroscopic approach to the Casimir friction force. *JETP Lett.* **2014**, *99*, 581–584.
81. Intravaia, F.; Behunun, R.O.; Dalvit, D.A.R. Quantum friction and fluctuation theorems. *Phys. Rev. A* **2014**, *89*, 050101(R), doi:10.1103/PhysRevA.89.050101.
82. Einstein, A.; Hopf, L. Statistische Untersuchung der Bewegung eines Resonators in einem Strahlungsfeld. *Ann. Phys.* **1910**, *338*, 1105–1115. (In German)
83. Einstein, A. Zur Quantentheorie der Strahlung. *Phys. Zeitsch.* **1917**, *18*, 121–128. (In German)

84. Mkrtchian, V.; Parsegian, V.A.; Podgornik, R.; Saslow, W.N. Universal Thermal Radiation Drag on Neutral Objects. *Phys. Rev. Lett.* **2003**, *91*, 220801, doi:10.1103/PhysRevLett.91.220801.
85. Høye, J.S.; Brevik, I.; Milton, K.A. Casimir friction between polarizable particle and half-space with radiation damping at zero temperature. *J. Phys. A* **2015**, *48*, 365004, doi:10.1088/1751-8113/48/36/365004.
86. Maghrebi, M.F.; Golestanian, R.; Kardar, M. Scattering approach to the dynamical Casimir effect. *Phys. Rev. D* **2013**, *87*, 025016, doi:10.1103/PhysRevD.87.025016.
87. Scheel, S.; Buhmann, S.Y. Casimir-Polder Forces on Moving Atoms. *Phys. Rev. A* **2009**, *80*, 042902, doi:10.1103/PhysRevA.80.042902.
88. Donaire, M.; Lambrecht, A. Velocity-dependent dipole forces on an excited atom. **2016**, *93*, 022701, doi:10.1103/PhysRevA.93.022701.
89. Volokitin, A.I. Blackbody friction force on a relativistic small neutral particle. *Phys. Rev. A* **2015**, *91*, 032505, doi:10.1103/PhysRevA.91.032505.
90. Dedkov, G.V.; Kyasov, A.A. Attraction Force, Frictional Torque, and Heating of a Spherical Particle Rotating in the Evanescent Electromagnetic Field of a Heated Surface. *Tech. Phys. Lett.* **2013**, *39*, 609–611.
91. Zhao, R.; Manjavacas, A.; Javier García de Abajo, F.; Pendry, J.B. Rotational Quantum Friction. *Phys. Rev. Lett.* **2012**, *109*, 123604, doi:10.1103/PhysRevLett.109.123604.